\documentclass[
amsmath, amssymb, 
aps, prapplied, reprint,
twocolumn,
superscriptaddress,
longbibliography,
floatfix,
showkeys,
xcolor
]{revtex4-1}

\usepackage[colorlinks, linkcolor = black, citecolor = black, filecolor = black, urlcolor = blue]{hyperref}

\usepackage{graphicx,braket,nicefrac}
\usepackage{siunitx,upgreek}
\usepackage{mathtools}

\newcommand{\sigx}[1]{\hat{\sigma}_x^{(#1)}}

\newcommand{\sigxSingle}{\hat{\sigma}_x}
\newcommand{\sigySingle}{\hat{\sigma}_y}

\newcommand{\creation}[1]{\hat{a}^\dagger_{#1}}
\newcommand{\annihilation}[1]{\hat{a}_{#1}}

\newcommand{\gam}[2]{\gamma_{#2}^{(#1)}}
\newcommand{\f}[2]{f_{#2}^{(#1)}}
\newcommand{\LD}[2]{\eta_{#2}^{(#1)}}

\newcommand{\eqdef}{=\vcentcolon}

\newcommand{\alphaScaledSumSq}[1]{\sum_k\vert \f{1}{k} \alpha_k({#1})\vert^2}

\newcommand{\alphaExpPlusPlus}[1]{e^{- 4\sum_k  \vert (\f{1}{k} + \f{2}{k}) \alpha_k({#1}) \vert^2 (\bar{n}_k+1/2)}}
\newcommand{\alphaExpPlusMin}[1]{e^{- 4\sum_k  \vert (\f{1}{k} - \f{2}{k}) \alpha_k({#1}) \vert^2 (\bar{n}_k+1/2)}}
\newcommand{\alphaExpSingle}[2]{e^{- 4\sum_k  \vert \f{#2}{k}\alpha_k({#1}) \vert^2 (\bar{n}_k+1/2)}}
 
\newcommand{\E}[1]{\mathbb{E}\left[#1\right]}

\begin{document}
\title{Phase-modulated entangling gates robust to static and time-varying errors}
\author{Alistair R. Milne}
	\thanks{These three authors contributed equally to this work.}
	\affiliation{
	ARC Centre of Excellence for Engineered Quantum Systems, The University of Sydney, School of Physics, NSW, 2006, Australia }
\author{Claire L. Edmunds}
	\thanks{These three authors contributed equally to this work.}
\affiliation{
	ARC Centre of Excellence for Engineered Quantum Systems, The University of Sydney, School of Physics, NSW, 2006, Australia }
    \affiliation{Q-CTRL Pty Ltd, Sydney, NSW, 2006, Australia}
\author{Cornelius Hempel}
	\thanks{These three authors contributed equally to this work.}
\affiliation{
	ARC Centre of Excellence for Engineered Quantum Systems, The University of Sydney, School of Physics, NSW, 2006, Australia }
\author{Federico Roy}
\thanks{Current address: Fachrichtung Physik, Universit\"at des Saarlandes, 66123 Saarbr\"ucken, Germany}
	\affiliation{
	ARC Centre of Excellence for Engineered Quantum Systems, The University of Sydney, School of Physics, NSW, 2006, Australia }
\author{Sandeep Mavadia}
	\affiliation{
	ARC Centre of Excellence for Engineered Quantum Systems, The University of Sydney, School of Physics, NSW, 2006, Australia }
\author{Michael J. Biercuk}
\thanks{michael.biercuk@sydney.edu.au}
	\affiliation{
	ARC Centre of Excellence for Engineered Quantum Systems, The University of Sydney, School of Physics, NSW, 2006, Australia }
     \affiliation{Q-CTRL Pty Ltd, Sydney, NSW, 2006, Australia}

\date{\today}     

\begin{abstract}
Entangling operations are among the most important primitive gates employed in quantum computing and it is crucial to ensure high-fidelity implementations as systems are scaled up. We experimentally realize and characterize a simple scheme to minimize errors in entangling operations related to the residual excitation of mediating bosonic oscillator modes that both improves gate robustness and provides scaling benefits in larger systems. The technique employs discrete phase shifts in the control field driving the gate operation, determined either analytically or numerically, to ensure all modes are de-excited at arbitrary user-defined times. We demonstrate an average gate fidelity of 99.4(2)\% across a wide range of parameters in a system of $^{171}\text{Yb}^{+}$ trapped ion qubits, and observe a reduction of gate error in the presence of common experimental error sources. Our approach provides a unified framework to achieve robustness against both static and time-varying laser amplitude and frequency detuning errors.  We verify these capabilities through system-identification experiments revealing improvements in error-susceptibility achieved in phase-modulated gates.  
\end{abstract}

\maketitle

\section{Introduction}

The ability to perform robust, high fidelity entangling gates in multi-qubit systems is a key requirement for realizing scalable quantum information processing \cite{NielsenChuang}. 
In several hardware architectures, qubits are entangled through shared bosonic oscillator modes via an interaction that is moderated by an external driving field. The M{\o}lmer-S{\o}rensen (MS) gate~\cite{Sorensen:1999,Solano:1999,Sorensen:2000} and the $\sigma_z$-gate~\cite{Leibfried:2003} in trapped ions as well as the resonator-induced phase gate in superconducting circuits \cite{Cross:2015,Paik:2016, Song:2017} are of this type. In addition, interactions simultaneously employing multiple bosonic modes have been explored to improve gate fidelities~\cite{McKay:2015} and probe novel types of interactions~\cite{Sundaresan:2015} in superconducting circuits.

A major source of error for oscillator-mediated gates is residual qubit-oscillator entanglement at the end of the operation~\cite{Bermudez:2017}. This detrimental effect can arise due to the presence of quasi-static or time-varying noise on the driving field, slow drifts in experimental parameters such as the qubit and oscillator frequencies, or the presence of spectator modes that are not properly accounted for in the gate construction. In trapped ion systems, various schemes have been demonstrated that minimize this residual coupling~\cite{Hayes:2012,Haddadfarshi:2016,Manovitz:2017,Shapira:2018,Webb:2018}, with some also incorporating the ability to simultaneously decouple from multiple modes~\cite{Choi:2014,Green:2015,Steane:2014,Schafer:2018,Leung:2018a,Leung:2018b,Lu:2019,Grzesiak:2019,Landsman:2019}. Their common feature is a temporal modulation of the driving field, modifying the trajectories of the joint qubit-oscillator states in each oscillator's phase space.

In this work, we experimentally demonstrate a new class of phase-modulated ($\Phi\text{M}$) entangling gates using trapped ions in the presence of multi-mode motional spectra. Specifically, we implement an MS-type interaction and employ discrete phase shifts of the driving field to suppress dominant gate errors.  Using both an analytic scheme~\cite{Green:2015} and numerical optimization to calculate the required phase shifts, we experimentally validate that phase modulation permits motional mode decoupling for arbitrary laser frequencies in a way not otherwise achievable through the conventional gate construction~\cite{Sackett:2000}. We achieve an average two-qubit gate fidelity of 99.4(2)\% (including SPAM errors) across a wide range of laser detunings near a pair of motional modes, reducing errors by up to two orders of magnitude relative to the best unmodulated alternative. We also demonstrate that proper construction of the $\Phi\text{M}$ sequence provides the ability to systematically increase gate-robustness to static offsets in the laser detuning, as well as time-varying laser detuning and amplitude noise.  Experimental measurements are in good agreement with a new theoretical model developed in the filter function framework~\cite{Green:2013} to capture the influence of time-dependent noise. Finally, we study the scaling behavior of both the analytic and numerically derived phase-modulated gate constructions with system size, and demonstrate that the use of numerical optimization reduces scaling behavior from exponential to linear with mode number, providing a means to accommodate high-fidelity, time-optimized $\Phi\text{M}$ gate construction in large multi-ion registers.

\section{Physical setting} 
\subsection{Oscillator-mediated entangling gates}

In oscillator-mediated entangling gates, the application of an external driving field, typically a microwave or laser, produces a qubit-state-dependent displacement of the oscillator wave packet in phase space. Given a system of $N$ qubits and $M$ bosonic oscillator modes, this coupling is described by the following time-dependent Hamiltonian
\begin{equation}
	\hat{H}(t) = i\hbar\sum_{\mu=1}^{N}\hat{\sigma}_{s}^{(\mu)}\sum_{k=1}^{M}\left( \gamma_{k}^{(\mu)}(t)\hat{a}^{\dagger}_{k} - {\gamma_{k}^{(\mu){*}}(t)}\hat{a}_{k}\right),
    \label{eq:MS_H}
\end{equation}
where $\hat{\sigma}_{s}^{(\mu)}$ is the Pauli spin operator in the basis \mbox{$s \in \{x,y,z\}$} acting on the $\mu\textsuperscript{th}$ qubit, and $\hat{a}^{\dagger}_{k},\,\hat{a}_{k}$ are creation and annihilation operators acting on the $k\textsuperscript{th}$ oscillator mode. The complex-valued function \mbox{$\gamma_{k}^{(\mu)}(t)= \Omega f_k^{(\mu)}e^{-i\delta_k t}$} describes the coupling of the $\mu\textsuperscript{th}$ qubit and $k\textsuperscript{th}$ oscillator mode, where the effective coupling strength is given by the product between the strength of the driving field $\Omega$ and a hardware-specific factor, $f_k^{(\mu)}$. Here, $\delta_k$ is the angular frequency difference (detuning) of the driving field from the $k\textsuperscript{th}$ oscillator mode. Under the application of the driving field, the coupled system undergoes a unitary evolution~\cite{Roos:2008,Kirchmair:2009} including both a qubit-qubit entangling term and a qubit-state-dependent displacement~$\hat{D}$ of the oscillator modes in phase space.  The latter is central to our discussion and is described by
\begin{equation}
    \hat{D} =\exp\left\{\sum_{\mu=1}^N \hat{\sigma}_s^{(\mu)} \sum_{k=1}^M\left[f_k^{(\mu)}\alpha_k(t)\hat{a}^{\dagger}_{k} - {f_k^{(\mu)*}{\alpha_k^*}(t)}\hat{a}_{k} \right]\right\}.
\end{equation}
Due to the detuning $\delta_k$, the wave packets associated with joint qubit-oscillator states undergo circular phase space trajectories proportional to the coherent displacement \mbox{$\alpha_k(t) = \Omega\int_{0}^{t} dt' e^{-i\left[\delta_{k}t^\prime + \phi(t')\right]}$}. The $k\textsuperscript{th}$ mode trajectory returns to its starting point with a period of $2\pi/|\delta_k|$ and the enclosed area of all wave packet trajectories is commensurate with the accumulated qubit-qubit entangling phase. Here, $\phi(t)$ is the phase difference between the oscillator and the driving field, which we refer to as the coupling phase and modulate in our approach.

Successful completion of a qubit-qubit entangling operation requires the elimination of qubit-oscillator entanglement; for a gate of length $t=\tau_g$, this is achieved by satisfying the condition \mbox{$\alpha_k(\tau_g) = 0$}, for each mode $k$.  Modulation of the coupling phase $\phi(t)$  may be used to direct the phase space trajectories, returning each trajectory to the origin in a shorter time than the typical approach of ensuring that the gate time and drive detuning are related by an integer multiple for all modes, $\delta_k\tau_g = 2\pi j$, for $j \in \pm\{1,2,...\}$. The ability to actively steer these trajectories is particularly important with large mode numbers where the gate time would otherwise grow prohibitively long, and even allows us to ensure effective mode decoupling in the presence of time-dependent parameter fluctuations. The required phase modulations may be determined through analytic calculation~\cite{Green:2015} or numerical optimization, and we now outline the details of these two approaches.  

\begin{figure}
    \includegraphics[scale=1]{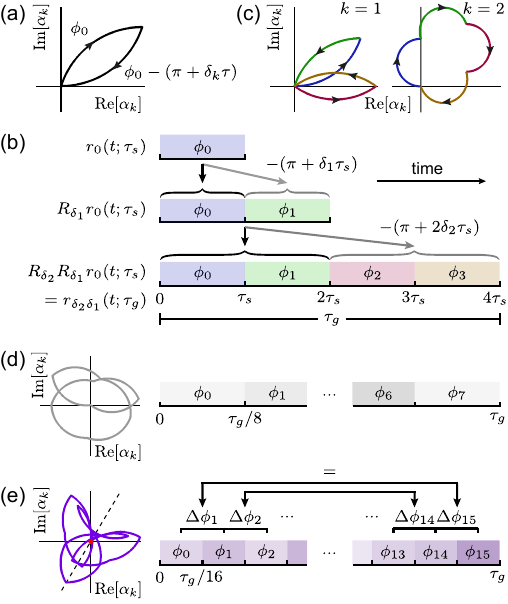}
    \caption{Construction of phase modulation sequences. (a) Schematic plot of $\alpha_k(t)$ for $0 \leq t \leq 2\tau$. Shifting the coupling phase, initially $\phi_0$, by an amount \mbox{$-(\pi +\delta_k\tau$)} at $t=\tau$ returns the oscillator trajectory to the origin at $t=2\tau$. (b) Construction of the $\Phi$M sequence $R_{\delta_2}R_{\delta_1}r_0(t;\tau_s)$ from the base sequence $r_0(t;\tau_s)$. Each application of $R_{\delta_k}$ produces a new sequence consisting of the original sequence (black arrow) followed by the entire original sequence phase-shifted by \mbox{$-(\pi +\delta_k\tau$)} (gray arrow), where $\tau$ is the duration of the original sequence. The example sequence closes the trajectories of two modes $k=1,\,2$, shown in (c), with detunings $\delta_1$,\,$\delta_2$ in four time segments of length $\tau_s = \tau_g/4$. Colors indicate the varying coupling phase $\phi_n$ in each time segment $t \in [n\tau_s,(n+1)\tau_s]$. (d) Example phase space trajectory (left) and schematic showing the construction (right) of a standard numerically optimized phase modulation sequence targeting two modes with $S = 8$ phase segments of $\tau_g/8$ duration each. (e) For the robust numerical sequence, the number of phase segments is doubled to $S = 16$, with each $\tau_g/16$ in length. The time-averaged position of the phase space trajectory (red dot) lies at the origin. This condition, combined with the constraint that $\Delta\phi_{n+1} = \Delta\phi_{S-(n+1)}$ results in a trajectory symmetric about a line through the origin (dashed line).}
\label{fig:concat}
 \end{figure}

\subsection{Calculation of phase modulation sequences}
The key to the analytic $\Phi\text{M}$ scheme is that for any time evolution of the $k$\textsuperscript{th} oscillator state over the interval $t \in [0,\tau]$, its phase space trajectory can be returned to the origin by repeating the same evolution over the interval $t \in [\tau,2\tau]$ with an overall shift of the coupling phase, $\phi(t)$, equal to $-(\pi+\delta_k\tau)$ (Fig.~\ref{fig:concat}(a)). Using the Heaviside function $\Theta(x)$, a segment $\tau$ of this evolution may be represented by \mbox{$r(x=t;\tau) = \Theta(t)\, \Theta(\tau-t)\, e^{-i\phi(t)}$}, modifying the qubit-oscillator coupling as \mbox{$\gamma_k^{(\mu)}(t) \rightarrow \gamma_k^{(\mu)}(t) r(t;\tau)$}. We define a family of operators $R_{\delta_k}$, parametrized by $\delta_k$ and acting on $r(x;\tau)$ as	\mbox{$R_{\delta_k}r(x=t;\tau) = r(t;\tau) + e^{i(\delta_k\tau+\pi)}r(t-\tau;\tau)$}. This captures a two-segment, piecewise-constant modulation sequence over the interval $t \in [0,2\tau]$, which returns the trajectory of mode $k$ to its initial state, yielding \mbox{$\alpha_k(2\tau) = 0$}. As illustrated in Fig.~\ref{fig:concat}(b, c), this process of phase-shifted concatenation may be repeated to construct sequences that close any number of oscillator trajectories in a desired gate time $\tau_g$. In order to decouple $M$ oscillators, the gate is divided into $2^M$ time segments of length \mbox{$\tau_s = \tau_g/2^M$} and the phase modulation sequence is constructed as $R_{\delta_M}...R_{\delta_1}r_0(t;\tau_s)$, where \mbox{$r_0(t;\tau_s) = \Theta(t)\Theta(\tau_s-t)$} is the `base' sequence for which we take $\phi(t) \equiv 0$. The short-hand notation \mbox{$r_{\delta_M...\delta_1}(t;\tau_g) \equiv R_{\delta_M}...R_{\delta_1}r_0(t;\tau_s)$} is used to refer to the full sequence of $2^M$ phase segments, where the phase in each segment may be calculated exactly using a closed-form expression~\cite{Green:2015}.

$\Phi\text{M}$ sequences that provide increased robustness to parameter fluctuations during the gate operation can be constructed by repeated application of the operator $R_{\delta_k}$ on the base sequence $r_0(t;\tau_s)$. The number of times the operator is applied determines the `order' of noise suppression associated with decoupling from mode $k$. A sequence that suppresses noise to order $(p+1)$ will achieve decoupling in the presence of noise that modifies the qubit-oscillator coupling via $\gamma_k^{(\mu)}(t) \rightarrow \gamma_k^{(\mu)}(t)\beta_k^{(p)}(t)$, where $\beta_k^{(p)}(t) = \sum_{j=0}^p\beta_{k,j}t^j$ is a $p\textsuperscript{th}$-order polynomial. For example, the sequence \mbox{$R_{\delta_2}R_{\delta_2}R_{\delta_1}r_0(t;\tau_s) = r_{\delta_2\delta_2\delta_1}(t;\tau_g)$} will decouple modes $k = 1$ and $k=2$, providing additional noise suppression to second order for mode $k=2$. These robust sequences result in mode trajectories that return to the origin repeatedly throughout the operation and, in the presence of noise, coherently average away deviations from the ideal oscillator trajectories. 

Numerical optimization can also be used to produce $\Phi\text{M}$ sequences that enable multi-mode decoupling.  This approach is designed to mitigate the unfavorable exponential scaling of the number of phase shifts with mode-number $M$ encountered in our analytic approach, trading closed-form solutions for the need to rely on numeric techniques (rather than a transparent physical argument) in finding them.  For a specified gate time $\tau_g$, set of drive-field detunings $\{\delta_k\}$, and number of phase segments $S$, the optimization procedure finds $\Phi\text{M}$ sequences that ensure all $M$ modes exhibit residual motional displacement below an arbitrarily defined threshold \mbox{$\sum_\mu \sum_k \left\vert f_k^{(\mu)} \alpha_k(\tau_g)\right\vert^2 \leq \epsilon$}. We find empirically that good solutions yielding $\epsilon\lesssim 10^{-4}$ are achievable using only a linear scaling in segment number $M$ with a small prefactor, $S=4M$ (see Fig.~\ref{fig:concat}(d)).

Robustness to fluctuations in experimental parameters may be realized in numerically optimized $\Phi\text{M}$ sequences by imposing the additional constraint that the time-averaged positions of the phase space trajectories for all modes lie at the origin, that is \mbox{$\alpha_{k, \textrm{avg}}(\tau_g) = \frac{1}{\tau_g}\int_0^{\tau_g}\alpha_k(t)dt = 0 $}. By further requiring that the trajectories be symmetric about half the gate time~\cite{Hayes:2012, Leung:2018a}, minimizing $\alpha_{k,\text{avg}}(\tau_g)$ is equivalent to minimizing $\alpha_k(\tau_g)$ and the optimization constraint becomes \mbox{$\sum_\mu \sum_k \left\vert f_k^{(\mu)} \alpha_{k, \textrm{avg}} (\tau_g)\right\vert^2 \leq \epsilon$}. The realization of symmetric phase space trajectories may be expressed as a condition on phase \emph{differences} between time segments in the two halves of the sequence, such that \mbox{$\Delta\phi_{n+1} = \Delta\phi_{S-(n+1)}$}. Here, \mbox{$\Delta\phi_n = (\phi_n - \phi_{n-1})$} and $\phi_n$ is the coupling phase in the time segment \mbox{$t \in [n\tau_g/S,(n+1)\tau_g/S]$}. In order to account for the additional symmetry constraint, the number of phase segments employed in the optimization is increased to $S=8M$, which ensures the optimizer routinely finds gate constructions satisfying the constraints. The difference in construction between the standard and robust numerically optimized $\Phi\text{M}$ sequences is illustrated in Fig.~\ref{fig:concat}(d,e).

\section{Results and Discussion}
\subsection{Experimental Setup}
We experimentally implement the schemes outlined above using a system of $^{171}\text{Yb}^{+}$ ions confined in a linear Paul trap (similar to \cite{Guggemos:2017}) with center-of-mass (COM) trap frequencies \mbox{$\omega_{x,y,z}/2\pi \approx (1.6,1.5,0.5)~\text{MHz}$}. Qubits are encoded in the $^2\mathrm{S}_{1/2}$ ground-state manifold where we associate the hyperfine states $\ket{F=0, m_F =0}\equiv\ket{0}$ and $\ket{F=1, m_F =0}\equiv\ket{1}$, split by 12.6~GHz, with qubit states $\ket{0}$ and $\ket{1}$ respectively. State initialization to $\ket{0}$ via optical pumping and state detection are performed using a laser resonant with the $^2\mathrm{S}_{1/2} - {^2\mathrm{P}_{1/2}}$ transition near 369.5~nm. 

A pulsed laser near 355~nm is used to drive stimulated Raman transitions between the qubit states~\cite{Hayes:2010} via two orthogonal laser beams in a geometry where they only couple to the $x,y$ radial motional modes of the trapped ions. To implement entangling gates, a two-tone radio-frequency signal produced by an arbitrary waveform generator is applied to an acousto-optic modulator controlling one of these beams. This produces a bichromatic light field that off-resonantly drives the red and blue sideband transitions, creating the state-dependent force used in the gate. Modulation of the coupling phase $\phi(t)$ is achieved by adjusting the phase difference between the red and blue frequency components, \mbox{$\phi(t) = \left[\phi_b(t) - \phi_r(t)\right]/2$}. The maximum achievable gate Rabi frequency is \mbox{$\Omega = 2\pi \times 40$~kHz}, limited by the available optical power.  

\subsection{Motional mode decoupling}

\begin{figure}[t]
    \includegraphics[scale=1]{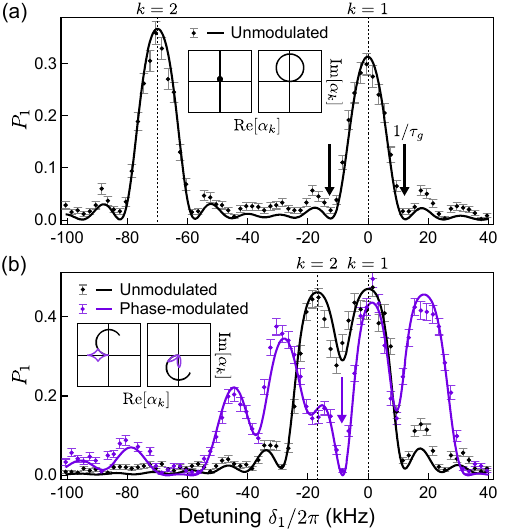}
    \caption{Motional mode decoupling in the single ion case. Population $P_1$ as a function of laser detuning scanned about two radial modes, $k=1$ and $k=2$ (dashed vertical lines). Solid lines are fits to the data, with each mode's initial average phonon number after sideband cooling ($\bar{n}_k$) a free parameter. (a) Mode frequencies split far apart. Arrows indicate detunings at which $\delta_1 = \pm2\pi/\tau_g$ and the inset shows corresponding phase space trajectories for each mode, with $\bar{n}_{1,\,2} = (0.2,0.4)$. (b) Mode frequencies closely spaced. The phase modulation sequence $r_{\delta_2\delta_1}(t;\tau_g)$ (purple) achieves decoupling at an arbitrary detuning $\delta_1/2\pi = -8.5$ kHz (arrow), with phase segments $\phi_{0,1,2,3} \approx (0,1.34,0.343,1.68)\, \pi$. The inset compares unmodulated and $\Phi\text{M}$ trajectories. Fits give $\bar{n}_{1,\,2} = (1.2, 1.2)$ for the unmodulated and $\bar{n}_{1,\,2} = (2.5, 1)$ for the $\Phi\text{M}$ data. }
    \label{fig:singleion}
\end{figure}

We begin by demonstrating the ability to arbitrarily decouple multiple motional modes using the analytic $\Phi\text{M}$ scheme. A single qubit is prepared in state $\ket{0}$ and the bichromatic Raman fields are applied for \mbox{$\tau_g=80~\upmu$s}. The Raman beams' frequency difference is scanned over a range including two radial modes. Here, the application of the state-dependent force produces a purely qubit-oscillator interaction and any residual mutual coupling at the conclusion of the operation will result in $P_{1}>0$, where population $P_{i}$ is the probability of $i$ ions being projected into state $\ket{1}$. 

In Fig.~\ref{fig:singleion}(a), we tune the modes to have a frequency splitting sufficiently large that the predominant interaction is with only a single mode. In this configuration, complete decoupling is achieved for the detunings $\delta_1 = \pm 2\pi/\tau_g$, indicated by $P_{1}$ dropping to zero symmetrically about $\delta_1 = 0$ (similarly about $\delta_1 = -70$ kHz, corresponding to $\delta_2 = 0$). In Fig.~\ref{fig:singleion}(b), the mode splitting has been decreased via electrostatic tuning of the trap potential such that both modes will become excited when the laser is detuned close to either, illustrating the problem of mode crowding typically experienced in larger systems. For an unmodulated Raman drive, the black data in Fig.~\ref{fig:singleion}(b) show a large value of $P_1$ at intermediate detunings (-17~to~0~kHz), where decoupling was previously achievable for mode $k=1$. By contrast, we may drive $P_{1}$ to zero at an arbitrarily chosen detuning (arrow) using a four-segment $\Phi\text{M}$ sequence to decouple both modes (purple data). The resulting phase space trajectories at this detuning are illustrated in the insets to Fig.~\ref{fig:singleion}(b), showing how the modulation protocol steers both trajectories back towards the origin at the conclusion of the gate. We have achieved similar decoupling at a range of arbitrary detunings via appropriate construction of the $\Phi\text{M}$ sequence.  

\subsection{Flexibility in gate operation}

\begin{figure}[t]
	    \includegraphics[scale=1]{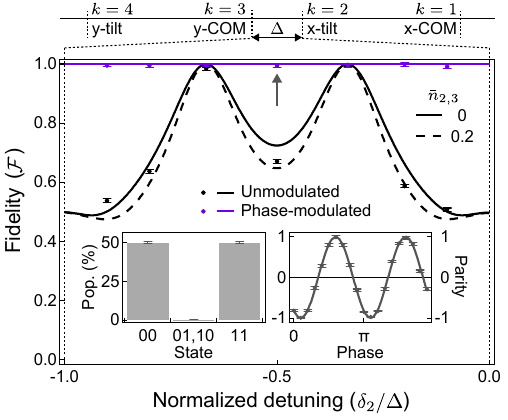}
    \caption{Maximum achievable gate fidelity as a function of detuning. Solid lines show theoretical predictions for initial phonon-numbers of $\bar{n}_{2,3} = 0$ and the dashed line show predictions for $\bar{n}_{2,3} = 0.2$. Different $\Phi\text{M}$ sequences are implemented over the detuning range, with $r_{\delta_2\delta_3}(t;\tau_g)$ used for $\delta_2/\Delta \geq -0.5$ and $r_{\delta_3\delta_2}(t;\tau_g)$ for $\delta_2/\Delta < -0.5$. The required Rabi frequency $\Omega$ ranges from $2\pi \times (24-26)$~kHz for the $\Phi\text{M}$ gates and $2\pi \times (19-26)$~kHz for the unmodulated gates. Error bars are derived from quantum projection noise on the state population estimates and a fit of the parity contrast. The inset shows the underlying data for the $\Phi$M gate at $\delta_2/\Delta = -0.5$ (arrow), for which $\mathcal{F}=99.4(5)\%$.}
    \label{fig:detflex}
\end{figure}

We now validate the impact of phase modulation in two-qubit MS entangling gates and demonstrate the flexibility it affords in choice of experimental parameters. For two qubits, there are four radial motional modes that may be excited by the Raman laser; we denote them from highest to lowest frequency as $k=1$ to $k=4$. Starting in state $\ket{00}$, we produce the entangled Bell-state \mbox{$(\ket{00} - i\ket{11})/\sqrt{2}$} by tuning the Raman laser fields to excite both the $x$-tilt ($k=2$) and $y$-COM ($k=3$) modes in our trap, separated by \mbox{$\Delta/2\pi \approx 10$}~kHz (Fig.~\ref{fig:detflex}). The remaining two modes are detuned by $\sim$~80 kHz, far enough to not be significantly excited. The gate time is chosen such that when the detuning from either mode is an integer multiple of $\Delta/3$, the spin and motion fully decouple, giving \mbox{$\tau_g = 2\pi \times 3/\Delta$} ($\sim 310$~$\upmu$s). Based on populations $P_0,P_1$ and $P_2$, the gate fidelity is estimated as \mbox{$\mathcal{F} = (P_0 + P_2)/2 + \pi_c/2$}. Here, $\pi_c$ is the parity contrast of the created Bell-state observed upon scanning the phase of an additional $\pi/2$-pulse after the gate. 

To demonstrate the flexibility of the analytic $\Phi\text{M}$ scheme, we vary the laser detuning over a range between the $y$-COM and $x$-tilt modes, optimizing the Rabi frequency $\Omega$ at each detuning to achieve a maximally entangling gate. Fig.~\ref{fig:detflex} compares the highest theoretically achievable fidelity (lines) for unmodulated and $\Phi$M-MS gates, along with experimental measurements (markers). For the unmodulated gate (black), maximum fidelity can only be achieved at two particular detunings where both mode trajectories naturally close. Elsewhere, the measured two-qubit gate fidelity drops to as low as 50\% due to strong residual mode excitation at the conclusion of the gate. In contrast, by implementing an appropriate $\Phi\text{M}$ gate, maximum fidelity can be ideally achieved for any detuning (purple line). For the $\Phi\text{M}$ data, we obtain an average experimental Bell-state fidelity of 99.4(2)\% across the range of detunings shown, without any form of SPAM subtraction. We estimate the contribution to the Bell state infidelity from imperfect state estimation to be 0.4(4)\%. 

\subsection{Suppressing static gate errors}

\begin{figure*}[b]
    \includegraphics[scale=1]{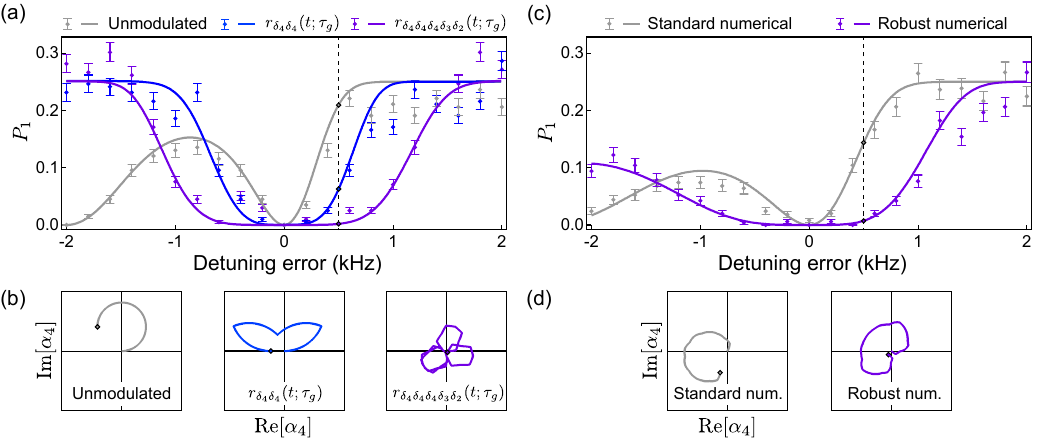}
    \caption{Robustness of $\Phi\text{M}$ gates to detuning offsets. (a) Motional excitation quantified by $P_1$ at the conclusion of a two-qubit entangling gate as a function of detuning offset magnitude. Solid lines are theory with $\bar{n}_4 = 0.1$. The gate time is fixed at $\tau_{g}=500$~$\upmu$s, with a target detuning of $-2$ kHz from mode 4. The unmodulated gate (gray) is compared to the $\Phi\text{M}$ sequences $r_{\delta_4\delta_4}(t;\tau_g)$ (blue) and $r_{\delta_4\delta_4\delta_4\delta_3\delta_2}(t;\tau_g)$ (purple), which provide second and third order noise suppression for mode 4, respectively. Rabi frequency $\Omega$ is scaled to enact a maximally entangling gate at zero detuning error and ranges from $2\pi \times (18-39)$~kHz. The scaling required for the third-order sequence also necessitates the decoupling of modes 2 and 3. (b) Phase space trajectories for mode 4 with a detuning error of +500 Hz, marked via the dashed line in (a). All plots are of equal scale. (c) $P_1$ at the conclusion of a two-qubit entangling gate for numerically optimized phase modulation sequences, constructed to sufficiently decouple from all four modes. Solid lines are theory with $\bar{n}_4 = 0.05$. The gate time is fixed at $\tau_{g}=400$~$\upmu$s with a target detuning of $-4$ kHz from mode 4. The standard numerically optimized gate (gray) with 16 phase segments is compared to a robust solution (purple) with 32 phase segments. Rabi frequencies are $2\pi \times 23$ kHz and $2\pi \times 28$ kHz for the standard and robust gates, respectively. (d) Corresponding phase space trajectories for mode 4 with a detuning error of +500 Hz, marked via the dashed line in (c).}
    \label{fig:detrobust}
\end{figure*}

An additional benefit of $\Phi\text{M}$ gates is the ability to incorporate robustness to imperfections in gate implementation. We explore this phenomenology by engineering static detuning offset errors during application of an entangling gate, and measuring $P_1$ as a proxy for gate infidelity associated with residual qubit-oscillator coupling. Such offsets are a common error and may arise due to slow drifts or incorrect calibration of the oscillator mode frequencies. 

In Fig.~\ref{fig:detrobust}(a) we illustrate this feature by performing two-ion $\Phi\text{M}$-MS gates constructed analytically with different orders of noise suppression for the target mode ($k=4$).  As the suppression order in the gate construction is increased, the range of detunings around zero for which $P_1 \approx 0$ broadens, demonstrating robustness for deviations up to $\pm1$~kHz from the target detuning value with third-order suppression. Data agree well with analytic theory (detailed in Appendix E) predicting the functional dependence of the measured $P_1$ on detuning.  Fig.~\ref{fig:detrobust}(b) demonstrates in the phase space picture how the process of phase-shifted sequence concatenation results in repeated decoupling of the mode throughout the gate, reducing residual excitation even for large detuning errors.

Similar benefits are also observed using robust numerically optimized $\Phi\text{M}$ sequences in Fig.~\ref{fig:detrobust}(c), where we compare the standard and robust numerical gate constructions, again in the presence of static detuning offset errors. In our experiments, the standard sequence does not provide robustness to such offsets, and we observe that $P_1$ as a function of the applied detuning error behaves similarly to the unmodulated gate in Fig.~\ref{fig:detrobust}(a). The symmetrization procedure and requirement that the time-averaged positions of the phase space trajectories are approximately zero for the robust gate solution reduces sensitivity to detuning errors, again indicated by the broadening of the dip in $P_1$.

\subsection{Reducing sensitivity to time-dependent noise}
\begin{figure*}[t!]
    \includegraphics[scale=1]{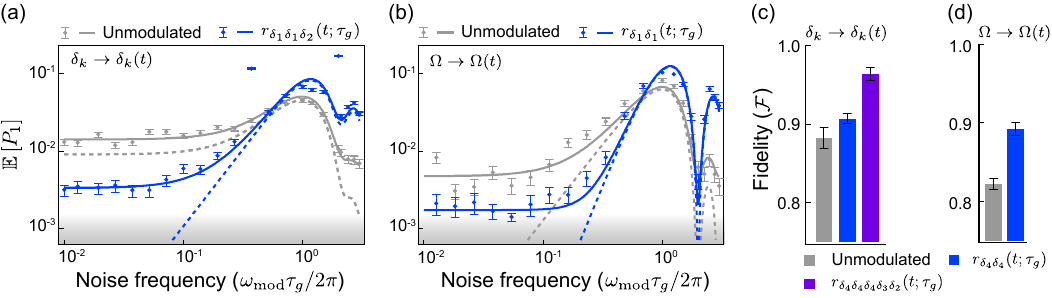}
    \caption{ 
Suppression of time-dependent noise. (a,b) Noise-averaged measurements of $P_1$ for a single ion plotted against applied (a) detuning and (b) amplitude noise frequency. Overlaid are the analytic filter function predictions with $\bar{n}_{1,2}=0.2$ (dashed lines) and the predictions with an added frequency-independent offset (solid lines), determined by the average of the six lowest-frequency data points. For (a), the added offsets are \mbox{$5.1\times10^{-3}$ (unmodulated)} and \mbox{$3.3\times10^{-3}$ ($\Phi\text{M}$)}. For (b), the added offsets are \mbox{$4.7\times10^{-3}$ (unmodulated)} and \mbox{$1.7\times10^{-3}$ ($\Phi\text{M}$)}. For (a) the modulation depth is $\beta=0.1$ with $\tau_g=$~300~$\upmu$s and for (b) $\beta = 0.29$ with $\tau_g=250~\upmu$s. The Rabi frequencies range from \mbox{$2\pi \times (14 - 34)$ kHz} and $\Omega$ is scaled in the $\Phi\text{M}$ operations to enclose the same phase space area as the equivalent unmodulated operation. Shading indicates the measurement floor. (c, d) Two-qubit gate fidelities with engineered (c) detuning and (d) amplitude noise. For both noise types, $\tau_g = 500$~$\upmu$s and $\Omega$ ranges from $2\pi \times (18 - 36)$~kHz. Here, the detuning noise is engineered by directly modulating the frequency of the motional modes via the application of a sinusoidally oscillating voltage to the DC trap electrodes. Measured gate fidelities for (c) are 86\%, 91\% and 96\%, respectively, under $\beta = 0.25$. For (d), the measured gate fidelities are 82\% and 87\%, under $\beta = 0.2$. Due to limitations on the experimentally achievable Rabi frequency in our setup (\mbox{$\Omega = 2\pi \times 40$~kHz}), only a second-order $\Phi\text{M}$ sequence is performed for the amplitude noise case shown in (d). }
	\label{fig:filter} 
\end{figure*}

Phase-modulated gates provide a third advantage in that they can provide robustness to time-varying experimental parameters in addition to static systematic offsets. This is useful in circumstances where parameters can drift or suffer from fluctuations due to, for example, electrical interference. In the following, we experimentally validate that $\Phi\text{M}$ gates may be used to provide robustness against errors induced by fluctuations in the motional mode frequencies and laser amplitude, which result in a time-dependent detuning offset and coupling-strength error, respectively. 

Gate effectiveness in suppressing time-dependent noise is conveniently captured using a filter function formalism~\cite{Green:2015,Soare:2014,Green:2012,Green:2013}, which describes the noise-susceptibility of an arbitrary control operation as a function of noise frequency. Modifications to the framework allow us to predict infidelity solely due to residual qubit-oscillator coupling in two-qubit entangling gates. For a given noise spectrum, $S_{\epsilon}(\omega)$, the decoupling error is inferred from the noise-averaged $P_1$ population, %
\begin{equation}
	\mathbb{E}\left[ P_{1} \right] \approx \frac{1}{2\pi}\int_{-\infty}^{\infty}d\omega S_{\epsilon} \left(\omega\right) F_{\epsilon}\left(\omega\right).
\end{equation}
Here, the filter function $F_{\epsilon}(\omega)$ expresses the susceptibility or ``admittance'' of the gate operation to a given noise source, with $\epsilon\in\{\Omega,\delta\}$ denoting laser amplitude or detuning noise, respectively. For laser amplitude noise, the analytic form of the filter function has been previously described~\cite{Green:2015}. The filter function for detuning noise represents an original contribution of this work and, and for the case of an operation performed on a single qubit, is given by
\begin{equation}
    F_{\delta}(\omega) =  \sum_k T_k\left\vert \Omega f_k^{(1)} \int_0^{\tau_g} dt e^{-i[(\delta_k - \omega)t + \phi(t)]} t \right\vert^2 
	\label{eq:decoupling_filter_function_main}.
\end{equation}
Here we have defined $T_k = 2\left(\bar{n}_k + 1/2\right)$, which incorporates the average phonon occupancy for each mode, $\bar{n}_k$. The filter function for a targeted entangling operation between a pair of qubits in an $N$-qubit system is described in Appendix F.  

To probe gate sensitivity under application of the two noise types described above with the highest possible measurement fidelity, we perform a gate-equivalent operation on a single ion in the presence of engineered noise in the respective quadrature. The gate detuning is set such that the motional interaction predominantly occurs with a single mode, and the decoupling condition is met ($\delta_1 = 2\pi/\tau_g$). We experimentally implement a system-identification procedure~\cite{Soare:2014} in which a single-frequency `noise' modulation is applied and produces an effective spectrum \mbox{$S(\omega_{\text{mod}}) \propto \delta(\omega-\omega_{\text{mod}}) + \delta(\omega+\omega_{\text{mod}})$}. A measurement of $\mathbb{E}\left[P_1\right]$ under this phase-averaged noise spectrum then gives a direct probe of the filter function at a single frequency, $\omega_{\text{mod}}$. We then vary $\omega_{\text{mod}}$ and measure $\mathbb{E}\left[P_1\right]$ at each value, effectively reconstructing the frequency-dependent filter function of the underlying gate operation. This approach is possible as the filter function Eq.~\eqref{eq:decoupling_filter_function_main} is only dependent on the residual qubit-oscillator coupling and independent of any entangling phase that would be acquired with two or more ions.

We engineer detuning noise via frequency-modulation of the two-tone RF signal producing the bichromatic gate beam; this has an effect on the gate interaction equivalent to fluctuating motional mode frequencies and modifies the laser detuning as \mbox{$\delta_k \rightarrow \delta_k(1+\beta\sin(\omega_{\text{mod}}t+ \phi_{\text{mod}}))$}, where $\beta$ quantifies the strength of the modulation, $\omega_{\text{mod}}$ its frequency, and $\phi_{\text{mod}}$ its phase. In a similar manner, laser amplitude noise is engineered by a direct modulation of the overall laser intensity via an acousto-optic modulator, yielding \mbox{$\Omega \rightarrow \Omega(1+\beta\sin(\omega_{\text{mod}}t+ \phi_{\text{mod}}))$}. 

The noise-suppressing properties of $\Phi\text{M}$ gates are experimentally validated in Fig.~\ref{fig:filter}(a,b).  In these experiments, we perform frequency-dependent system identification on an unmodulated and an analytic $\Phi$M sequence constructed to exhibit second-order suppression of both detuning and laser-amplitude noise. Experimentally, we observe that the $\Phi\text{M}$ gate (blue) exhibits lower measured error across the range of applied noise frequencies until a crossover is reached near the inverse gate time $\omega_{\textrm{mod}}\tau_g/2\pi \approx 1$. This behavior indicates a trade-off between low frequency error suppression and sensitivity to noise near the inverse gate time, consistent with observations for single-qubit operations \cite{Soare:2014}.

We find good agreement between experimental measurements and an empirical model combining the prediction for gate error from the filter function with a frequency-independent error offset extracted from measurement (solid lines). For both noise types, the filter function predictions for $\Phi$M~gates (dashed lines) show decreased noise sensitivity in the regime $\omega_{\textrm{mod}}\tau_g/2\pi < 1$, captured by an enhanced slope on a log-log plot. In the case of detuning noise shown in (a), the base filter function prediction for the unmodulated gate exhibits broadband sensitivity to noise, manifested in the saturation of $\mathbb{E}\left[P_{1}\right]$ towards lower noise frequencies. For the unmodulated gate exposed to amplitude noise as shown in (b), the filter function prediction drops towards zero in the low frequency regime, as a quasi-static error will simply result in a scaling of the Rabi frequency and will not affect the closure of the phase space trajectories. 

The error offset employed in our empirical model may arise in the experiment due to uncompensated noise from some uncontrolled source, or potentially from higher-order modulation-frequency-dependent terms in the filter function~\cite{Green:2012, Paz-Silva:2014} (\emph{i.e.} higher-order sensitivity to the applied single-frequency noise source).  In error-suppressing gates we have previously observed that when studying agreement between measurements and filter function predictions, higher-order filter function contributions grow in importance when leading-order error terms are cancelled by a compensating pulse~\cite{Soare:2014}. However, in these measurements we find larger error offsets associated with unmodulated gates, consistent with the presence of an intrinsic noise process to which the unmodulated gate is more susceptible.  

We next demonstrate that $\Phi\text{M}$ gates provide error-suppression in two-qubit entangling operations subjected to time-dependent noise, measuring the full gate fidelity. As in the system-identification routine above, we apply a single-frequency noise modulation, with a normalized frequency \mbox{$\omega_{\textrm{mod}}\tau_g/2\pi = 0.1$}. Here, the noise modulation is sufficiently strong to reduce the unmodulated (no $\Phi\text{M}$) gate fidelity below 90\%.  This fidelity-loss can be substantially recovered though use of $\Phi\text{M}$; replacing an unmodulated gate with a $\Phi\text{M}$ construction (Fig.~\ref{fig:filter}(c,d)), we observe an increase in gate fidelity as the order of error suppression is increased for both noise types.

\subsection{Scaling to larger systems}

When performing $\Phi \text{M}$ gates in systems with a large number of oscillator modes, one needs to consider the associated growth in the number of phase segments required to achieve mode decoupling. In ion trap systems, every additional ion contributes one axial and two radial modes to the spectrum. Increasing the number of phase segments typically results in trajectories that enclose a smaller phase space area, necessitating either an increase in the Rabi frequency or gate time in order to accumulate the desired entangling phase. Ideally, both of these quantities should be minimized, reducing sensitivity to error sources such as photon scattering \cite{Ozeri:2007} or, in the case of increased gate time, motional heating and motional dephasing. To explore the relevant scaling behavior for both analytic and numerically optimized constructions, we consider the case of a fixed maximum Rabi frequency and calculate the shortest achievable gate time realized in the two approaches, as a function of ion number in a linear chain.

\begin{figure}[t]
    \includegraphics[scale=1]{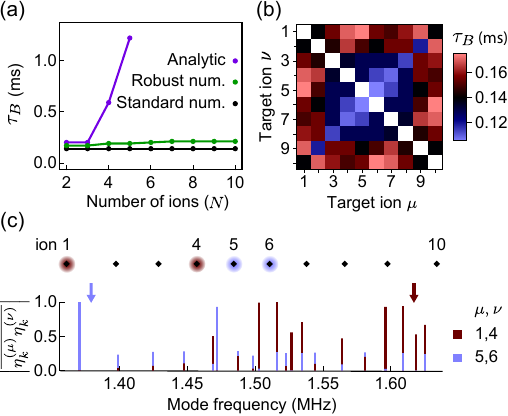}
    \caption{ 
Scaling to larger qubit systems. (a) Shortest achievable gate time ($\tau_B$) as a function of the number of qubits, decoupling from all $M=2N$ radial modes. The number of phase shifts for each $\Phi\text{M}$ approach is fixed at: $S=2^M$ (analytic), $S=4M$ (standard numerical) and $S=8M$ (robust numerical). The mode spectrum is calculated using COM trapping frequencies of \mbox{$\omega_{x,y,z}/2\pi = (1.62,1.54,0.15)$ MHz} with the maximum Rabi frequency limited to \mbox{$\Omega = 2\pi \times 100$ kHz}. (b) Shortest achievable gate time ($\tau_B$) represented as a colorscale for standard numerically optimized entangling gates between different target pairs in an $N=10$ ion chain, decoupling from the radial modes. Black represents the median gate time, red indicates slower gates and blue indicates faster. (c) Schematic depiction of the equilibrium positions for a $N=10$ ion chain and corresponding radial mode spectrum. Vertical bars positioned at the frequency of each mode $k$ indicate the value of \mbox{$|\overline{\eta_k^{(\mu)}\eta_k^{(\nu)}}|$}, which is $|\eta_k^{(\mu)}\eta_k^{(\nu)}|$ normalized to the maximum value of $|\eta_k^{(\mu)}\eta_k^{(\nu)}|$ for all $k$. Arrows indicate the detunings corresponding to $\tau_B$ for two target pairs: 1,4 (red) and 5,6 (blue).     }
	\label{fig:scaling} 
\end{figure}

As a concrete example, we consider an entangling operation between the two outermost adjacent ions in an $N$-ion chain of $^{171}$Yb$^+$ ions. For each $N$, we fix $\Omega = 2\pi\times 100$~kHz and perform a discretized search over detuning and gate time, choosing the $\Phi\text{M}$ sequence that results in a maximally entangled pair in the shortest time $\tau_B$. The $\Phi\text{M}$ sequences are constructed to decouple from all $M=2N$ radial motional modes such that $\epsilon \lesssim 10^{-4}$. As shown in Fig.~\ref{fig:scaling}(a), with increasing $N$, the numerically optimized sequences scale more favourably than those calculated analytically. For the standard numerically optimized sequences, we find the shortest gate time is constant ($140$~$\upmu$s) for up to $N=10$ ions. In the robust case, $\tau_B$ is slower than the standard sequences by a small offset and grows gradually with ion number from 170~$\upmu$s~($N=2$) to 210~$\upmu$s~($N=10$). The rapid increase in $\tau_B$ for the analytic gate construction, combined with the simultaneous exponential growth in the required number of phase segments ($2^M$), ultimately render it less suitable for larger systems than the numerically derived alternative. Also, despite the fact that a specific analytic $\Phi\text{M}$ sequence may be calculated in closed form, the process of choosing the best analytic construction from all possible permutations of \mbox{$r_{\delta_M...\delta_1}(t;\tau_g)$} sequences can add computational complexity, as the ordering of mode closure represents an additional degree of freedom that impacts gate time. To avoid this computational overhead, for the analytic data shown in Fig.~\ref{fig:scaling}(a), we consider only a single permutation corresponding to mode closure in order from largest to smallest detuning. This ordering tends to minimize the error-contribution from far-detuned modes by keeping their trajectories close to the origin, while also producing trajectories for the strongly excited modes that result in a larger accumulated entangling phase.  

These observations also hold for entangling operations performed on arbitrary ion pairs within a larger chain, where spatial variation in ion-motion coupling across the chain provides an additional consideration in gate construction. In Fig.~\ref{fig:scaling}(b), we investigate the dependence of $\tau_B$ on the choice of target ion pair within an $N=10$ ion crystal for standard numerically optimized gates, again requiring decoupling from just the radial motional modes. Assuming a $2\pi\times100$~kHz maximum Rabi frequency, $\tau_B$ ranges from 120~$\upmu$s to 175~$\upmu$s. This variation highlights the fact that the rate at which entanglement is accumulated between ions $\mu,\nu$ will depend on how strongly the ions are coupled to the dominant excited modes, as quantified by the Lamb-Dicke parameters \mbox{$\eta_k^{(\mu)},\eta_k^{(\nu)}$}. For example, a particular mode may involve a large displacement of the outer ions, while the central ions remain almost stationary. This mode would therefore be unsuitable for use in entangling gates attempting to induce coupling between central and outer ions. Such variability in the coupling strength across a chain is shown for two different ion pairs in Fig.~\ref{fig:scaling}(c), where we plot the magnitude of the Lamb-Dicke parameters across a complex multi-mode spectrum.

Considering this additional complexity, the flexibility provided by $\Phi\text{M}$ sequences to set the detuning arbitrarily within the mode spectrum enables the leveraging of entangling-phase contributions from multiple modes where $\left|\eta_k^{(\mu)}\eta_k^{(\nu)}\right|$ is greatest for the given ion pair. As such, the use of $\Phi\text{M}$ gates also offers speed advantages relative to conventional techniques where the detuning would simply be fixed close to a COM mode and the gate speed limited by the requirement that the other modes are not significantly excited. Using phase modulation, we are free to set the detuning arbitrarily for gates implemented between different target ion pairs; the arrows in  Fig.~\ref{fig:scaling}(c) show the detuning values corresponding to $\tau_{B}$ for gates between two different ion pairs. Without phase modulation, the detuning could not be set arbitrarily while still achieving a high-fidelity gate (see Fig.~\ref{fig:detflex}).

\section{Conclusion}
We have demonstrated that phase modulation provides a robust and flexible framework for performing high-fidelity entangling gates in qubit-oscillator systems, validated using trapped ions. We have implemented two-qubit entangling gates with an average fidelity of $99.4(2)\%$, achieving maximum fidelity for arbitrary laser detunings, including settings where unmodulated gates cannot be achieved with high fidelity. In addition, we have shown that the $\Phi\text{M}$ framework provides robustness to static and time-dependent errors in the laser amplitude and gate detuning, captured through a new theoretical model in the filter function framework.

The $\Phi\text{M}$ approach to constructing oscillator-mediated entangling gates gates holds several practical advantages relative to alternative modulation approaches \cite{Webb:2018, Choi:2014}. First, the amplitude and frequency of the driving field remain fixed throughout the gate operation, meaning experimental -– often duty-cycle-dependent –- nonlinearities and the effect of time-dependent AC Stark shifts need not be considered. Next, the $\Phi\text{M}$ technique is also readily extensible to multi-qubit entangling gates in larger systems, where we have shown that numerical optimization may be utilized to exponentially reduce the number of phase shifts required to decouple from an increasing number of modes, and flexibility in gate detuning can be used to extract speedups in gate implementation. We hope that the $\Phi\text{M}$ techniques demonstrated here may be employed beyond ion trap systems to improve gate fidelity in other architectures that utilize oscillator-mediated operations. 

\section*{Acknowledgments}
We acknowledge Todd Green for insightful discussions. This work was supported by the Office of the Director of National Intelligence (ODNI), Intelligence Advanced Research Projects Activity (IARPA), via the U.S. Army Research Office Grant No. W911NF-16-1-0070, and the ARC Centre for Engineered Quantum Systems CE110001013. We further acknowledge a private grant from H. \& A. Harley.

\section*{Appendix A: Maximum-likelihood procedure for state estimation}

Hyperfine qubits such as $^{171}\text{Yb}^+$ are susceptible to leakage between the qubit states from off-resonant excitations during the laser-induced fluorescence measurement. We employ a maximum-likelihood state estimation procedure~\cite{Webb:2018}, which reduces measurement error by accounting for the effect of state decay based on independent calibrations. In a given two-qubit experiment, we would like to determine the resultant state populations $P_i$, which are the probabilities for measuring $i$ ions in the $\ket{1}$ state. To do this, the ions are illuminated with a 369 nm laser, projecting each qubit into either the $\ket{1}$ (bright) or $\ket{0}$ (dark) state. The experiment is repeated $n$ times and the number of photon counts measured on an avalanche photodiode (APD) for each repetition is recorded. The resultant count rates are plotted in a histogram and by setting two count rate thresholds $c_1$ and $c_2$ between the resulting distributions, each repetition is assigned a outcome of `$i$ ions bright', where $i\in\{0,1,2\}$. We denote the number of repetitions assigned to each outcome as $x_i$, with \mbox{$x_0 + x_1 + x_2 = n$}. Due to state decays, there will be an overlap in the count rate distributions, meaning the true probabilities $P_i$ are not simply the proportion of repetitions assigned to outcome $i$. Instead, we define a linear map relating the measured probabilities $P_i^\prime = x_i/n$ to the true probabilities $P_i$, which takes the form \mbox{$P_i^\prime = \sum_j P(i\vert j)P_j$}. Here, $P(i\vert j)$ is the probability of classifying a repetition as `$i$ ions bright' given the ions were prepared in the state `$j$ ions bright'. The probabilities $P(i\vert 0)$ and $P(i\vert 2)$ are obtained by preparing and measuring the states $\ket{00}$ and $\ket{11}$, respectively, in calibration experiments. Without the ability to individually address ions, it is not possible to prepare the state $\ket{01}$ or $\ket{10}$, hence to determine the probabilites $P(i\vert 1)$, we assume $P(2\vert 1) = 0$ (which is a fair assumption given our detection duration and laser powers) and utilize a single ion to obtain $P(0\vert 1)$ and $P(1\vert 1)$. For a given set of repetitions, we compute the log-likelihood function $f(P_1,P_2)$ for discretized values of $P_1,P_2$ between 0 and 1. The values of $P_1,P_2$ that maximize $f(P_1,P_2)$ are the most probable given the data~\cite{Webb:2018}.

\begin{widetext}
\begin{equation}
    f(P_1,P_2) = \log\left(\frac{(n+1)(n+2)n!P_1^{\prime}(P_1,P_2)^{x_1}P_2^{\prime}(P_1,P_2)^{x_2}(1-P_1^{\prime}(P_1,P_2)-P_2^{\prime}(P_1,P_2))^{n-x_1-x_2}}{x_1!x_2!(n-x_1-x_2)!}\right)
    \label{eq:max_likelihood}
\end{equation}
\end{widetext}

Note that Eq.~\eqref{eq:max_likelihood} only depends on populations $P_1$ and $P_2$ as normalization always enables $P_0$ to be expressed as \mbox{$1-P_1-P_2$}. In order to make $f(P_1,P_2)$ computable, Stirling's approximation ($\log(n!) \approx n\log(n) - n$) must be used as terms such as $n!$ diverge too rapidly to be calculated for large $n$.

\section*{Appendix B: Bell state measurement fidelity}

To calculate how the measured Bell state fidelity is affected by imperfect state estimation via the maximum likelihood (ML) scheme, we follow an approach similar to the one outlined in~\cite{Ballance:2014}. By preparing and measuring known input states, we again construct a linear map relating the true populations $P_i$ to the population outcomes determined by the ML procedure $P_i^{''}$. In this case, the values $P(i|j)$ represent the probability that the ML procedure assigns some population `$i$ ions bright' to the measurement outcome, given that a state with population `$j$ ions bright' is prepared. \\

As described in the main text, the application of the MS gate results in the Bell state \mbox{$(\ket{00}+i\ket{11})/\sqrt{2}$}, the fidelity of which may be expressed as

\begin{equation}
\mathcal{F} = \frac{P_0+P_2}{2} + \frac{\pi_c}{2}.
\label{eq:bare_fidelity}
\end{equation}
Here, $\pi_c$ is parity contrast of the resultant state, measured by varying the phase of an additional $\pi/2$ analysis pulse at the conclusion of the gate. Assuming a Bell state with perfect fidelity, $\pi_c$ may be expressed as half the difference in parity between two states \mbox{$\ket{E} = (\ket{00}-i\ket{11})/\sqrt{2}$} and $\ket{O} = (\ket{01}+i\ket{10})/\sqrt{2}$ with parities $\mathcal{P}=1$ and $\mathcal{P}=-1$, respectively. Hence the expression for the fidelity (Eq.~\ref{eq:bare_fidelity}) becomes
\begin{equation}
\begin{aligned}
\mathcal{F} &= \frac{P_{0,E}+P_{2,E}}{2} \\ 
&+ \frac{(P_{0,E}+P_{2,E}-P_{1,E})-(P_{0,O}+P_{2,O}-P_{1,O})}{4},
\label{eq:fidelity}
\end{aligned}
\end{equation}
where $P_{i,k}$ indicate the populations for the ideal Bell states $\ket{k}$, with $k \in \{E,O\}$. \\

We model the effect of imperfect state estimation by substituting the true populations $P_{i,k}$ in Eq.~\ref{eq:fidelity} with the measured populations $P_{i,k}^{''}$, which are related to the true populations by the linear map. As an example, we may calculate the measured populations $P_{i,E}^{''}$ as

\begin{equation}
\begin{aligned}
\begin{pmatrix}
P_{0,E}^{''} \\ 
P_{1,E}^{''} \\ 
P_{2,E}^{''} \\
\end{pmatrix} &= 
\begin{pmatrix}
P(0|0) & P(0|1) & P(0|2) \\
P(1|0) & P(1|1) & P(1|2) \\
P(2|0) & P(2|1) & P(2|2) \\
\end{pmatrix}
\begin{pmatrix}
1/2 \\ 
0 \\ 
1/2 \\
\end{pmatrix} \\
&= \begin{pmatrix}
P(0|0) + P(0|2) \\
P(1|0) + P(1|2) \\
P(2|0) + P(2|2) \\
\end{pmatrix}.
\end{aligned}
\end{equation}
Re-expressing Eq.~\ref{eq:fidelity} in terms of the probabilities $P(i|j)$, we arrive at an expression for the Bell state fidelity incorporating imperfect state estimation.

\begin{equation}
\begin{aligned}
    \mathcal{F} &= 1-\frac{1}{2}\left[P(0|1)+P(1|2)+P(1|0)+P(2|1)\right] \\
    &= 1- \epsilon
    \end{aligned}
\end{equation}
Hence the contribution to the measured infidelity due to state estimation is given by

\begin{equation}
    1-\mathcal{F} = \epsilon = \frac{1}{2}\left[P(0|1)+P(1|2)+P(1|0)+P(2|1)\right].
    \label{eq:infidelity}
\end{equation}

The matrix below shows typical values of $P(i|j)$, which inserted into Eq.~\ref{eq:infidelity} give an error contribution of $\epsilon\approx$~0.4(4)\%. The quoted uncertainties are the standard deviation over multiple calibration measurements. 

\begin{align}
P(i|j) = \begin{pmatrix}
0.997(4) & 0.001(3) & 0.0002(6) \\
0.002(4) & 0.997(3) & 0.003(6) \\
0.0002(4) & 0.001(1) & 0.996(6) \\
\end{pmatrix}
\end{align}

\section*{Appendix C: M{\o}lmer-S{\o}rensen Time Evolution}

The Hamiltonian \eqref{eq:MS_H} describes a system of $N$ qubits coupled to $M$ oscillator modes via an external driving field. It results in the unitary evolution
\begin{align}
	\hat{U}(t) = \text{exp}\left\{ i\sum_{\mu,\nu=1}^{N} \varphi_{\mu\nu}(t) \sigx{\mu}\sigx{\nu} + \sum_{\mu=1}^{N} \sigx{\mu} \hat{B}_{\mu}(t) \right\},
    \label{eq:MS_U}
\end{align}
which is obtained from the Magnus expansion, where all higher order terms are identically zero.
This evolution has two key components: for \mbox{$N>1$} qubits, the first term describes pairwise entanglement between qubits $\mu, \, \nu$, captured by the phase
\begin{equation}
	\label{eq:entangling_phase}
    \varphi_{\mu\nu}(t) =  \textrm{Im}\left[\sum_{k=1}^M \int_0^t d{t_1} \int_0^{t_1} d{t_2} \gam{\mu}{k}(t_1) {\gam{\nu}{k}}^*(t_2) \right],
\end{equation}
and the second term describes a qubit-state-dependent displacement of the oscillator modes, via the displacement operators $\hat{D}_k$,
%
\begin{align}
    \label{eq:displacement_operator}
    &\exp\left\{\sum_{\mu=1}^{N} \sigx{\mu} \hat{B}_{\mu}(t) \right\} \\ &= \textrm{exp} \left\{\sum_{\mu=1}^N \sigx{\mu} \sum_{k=1}^M\left( f_{k}^{(\mu)}\alpha_k(t) \creation{k} - {f_{k}^{(\mu)}}^* {\alpha_k(t)}^* \annihilation{k} \right)\right\} \nonumber \\
    &= \prod_{k=1}^M \hat{D}_k \left(  \sum_{\mu=1}^N \sigx{\mu} f_{k}^{(\mu)}\alpha_k(t) \right).
\end{align}

The $N$-qubit system can be described by \mbox{$2^N$} eigenstates in the $[\sigxSingle]^{\otimes N}$-basis. Under the action of the displacement operator, the wave packet associated with the $k^\textrm{th}$ oscillator splits, as each component becomes entangled with one of the qubit eigenstates and is coherently displaced along a trajectory in phase space proportional to 

\begin{equation}
\alpha_{k}(t) = \Omega\int_{0}^{t} dt^{'} e^{-i\left[\delta_{k}t^{\prime}+\phi(t^{\prime})\right]}.
\end{equation}

In our system, the state-dependent displacement of the collective ion motion is produced by a bichromatic laser field, with frequency components tuned below (red) and above (blue) the bare qubit transition, offset from the motional mode frequencies by the detuning $\delta_k$. This detuning is defined to be \mbox{$\delta_k = \Delta\omega_{b} - \omega_0 - \omega_k $}, where $\Delta\omega_b$ is the frequency difference between the two Raman beams for the blue component of the bichromat, $\omega_0$ is the qubit frequency, and $\omega_k$ the frequency of the $k^\textrm{th}$ motional mode. The qubit interaction basis $\hat{\sigma}_s$ is determined by the optical phases $\phi_r$ and $\phi_b$ of the red and blue components respectively, such that \mbox{$\hat{\sigma}_s = \cos{[\frac{\phi_r+\phi_b}{2}]}\sigxSingle + \sin{[\frac{\phi_r+\phi_b}{2}]}\sigySingle$}. By setting $\phi_r = -\phi_b$, we fix the interaction basis to be $\hat{\sigma}_s = \sigxSingle$.

In the context of trapped ions, the hardware-specific coupling factor $f_k^{(\mu)}$ that relates the coupling of the $k$\textsuperscript{th} motional mode to the $\mu$\textsuperscript{th} ion is captured by the Lamb-Dicke parameter $\LD{\mu}{k}$,
\begin{align}
	\label{eq:fkmu}
    f_k^{(\mu)} =  \frac{-i\LD{\mu}{k}}{2} = \frac{-i}{2} b_{k}^{(\mu)} \Delta k \cos{\theta} \sqrt{ \frac{\hbar}{2 m \omega_k} }.
\end{align}
The Lamb-Dicke parameter incorporates the relevant normal mode eigenvector element, $b_{k}^{(\mu)}$, as well as the overlap of the spatial orientation of the motional mode with the effective wavevector of the driving field~\cite{James:1998}. Here, $\Delta k$ is the net wavevector of the two Raman beams, $\theta$ is the angle between the wavevector and the mode orientation, $\omega_k$ is the angular frequency of the mode, and $m$ is the mass of a single ion.

\section*{Appendix D: Calculation of observables after M{\o}lmer-S{\o}rensen evolution}

We expand on the approach presented in \cite{Roos:2008}, \cite{Kirchmair:2009} to calculate analytic expressions for observable quantities after the application of the M{\o}lmer-S{\o}rensen time evolution operator \eqref{eq:MS_U}. For a two-qubit ($N=2$) system, we assume an initially separable qubit-oscillator state with the qubits initialized to $\ket{00}$ and each oscillator in a thermal state with mean phonon number $\bar{n}_k$. To calculate expressions for the expectation value of the populations, $P_i(t)$, we use the projection operators \mbox{$\hat{P}_{lm} = \ket{lm}\bra{lm}$}, \, \mbox{$l, m \in \{0, 1\}$}, to obtain

\begin{widetext}
\begin{subequations}
	\label{eq:two_ion_populations}
    \begin{align}
        P_0(t) &= \langle \hat{P}_{00}(t)\rangle
        = \frac{1}{8} \Big(2 + \alphaExpPlusPlus{t}
        + \alphaExpPlusMin{t} \nonumber \\
        &\hspace{75pt} + 4\cos[4\varphi(t)]\alphaExpSingle{t}{1} \Big)  \\
        P_1(t) &= \langle \hat{P}_{01}(t)\rangle + \langle \hat{P}_{10}(t)\rangle  \nonumber \\
        &= \frac{1}{4}\left( 2 - \alphaExpPlusPlus{t} - \alphaExpPlusMin{t} \right)  \\
        P_2(t) &= \langle \hat{P}_{11}(t)\rangle
        = \frac{1}{8} \Big(2 + \alphaExpPlusPlus{t} 
        + \alphaExpPlusMin{t} \nonumber \\
        &\hspace{75pt} - 4\cos[4\varphi(t)]\alphaExpSingle{t}{1} \Big).
    \end{align}
\end{subequations}
\end{widetext}

The expression for $P_1(t)$ is used to plot the theory lines in Fig.~\ref{fig:detrobust}. The expressions for $P_0(t)$ and $P_2(t)$ are incorporated in the prediction for the Bell State fidelity (Fig.~\ref{fig:detflex}), given by $\mathcal{F} = \left(P_0 + P_2\right)/2 + \pi_c/2$. The additional quantity, $\pi_c$ (the parity contrast), in the formula for the fidelity corresponds to the magnitude of the off-diagonal elements of the density matrix describing the final two-qubit electronic state. Combining these elements, the full expression used in Fig.~\ref{fig:detflex} for the predicted Bell State fidelity is   

\begin{widetext}
\begin{align}
	\mathcal{F} &= \frac{1}{8}\left(2 + \alphaExpPlusPlus{t} + \alphaExpPlusMin{t} \right. \nonumber \\
	&\hspace{25pt} + \textrm{Abs} \left[  \alphaExpPlusPlus{t} - \alphaExpPlusMin{t} \right. \nonumber \\ 
	&\hspace{55pt}\left.\left.-4i\alphaExpSingle{t}{1}\sin{[4\varphi(t)]} \right] \right).
    \label{eq:exp_bell_state_fidelity}  
\end{align}
\end{widetext}

In the case of a single ion initialized to $\ket{0}$, the expressions for the populations after MS time evolution are the following

\begin{subequations}
	\label{eq:one_ion_populations}
    \begin{align}
    P_0(t) &=  \frac{1}{2}\left( 1 + \alphaExpSingle{t}{1} \right) \\
    P_1(t) &=  \frac{1}{2}\left( 1 - \alphaExpSingle{t}{1} \right).
	\end{align}
\end{subequations}

\section*{Appendix E: Numerical optimization of $\Phi\text{M}$ sequences.}

The numerical optimization of $\Phi\text{M}$ M{\o}lmer-S{\o}rensen gates is performed utilizing MATLAB's inbuilt constrained optimization routine \textit{fmincon}. For the optimization, we consider a targeted entangling operation between two ions $\mu,\nu$ in an $N$ ion chain with $2N$ radial motional modes. The motional frequency spectrum is either numerically calculated using the desired $x$-COM and $y$-COM frequencies, or manually input from experimental measurements. We specify the number of phase segments $S$, drive field detunings $\{\delta_k\}$, gate time $\tau_g$ and maximum Rabi frequency $\Omega_{\text{max}}$. For these parameters, the optimization procedure finds $\Phi\text{M}$ sequences that maximize the acquired entangling phase between the two target ions $\varphi_{\mu\nu}(\tau_g)$. This optimization occurs subject to the constraint that the residual motional displacement remain below a threshold of $10^{-4}$, that is

\begin{equation}
    \sum_{i = \mu,\nu} \sum_{k=1}^{M} \left\vert\frac{1}{2} \eta_k^{(i)}\alpha_k(\tau_g)\right\vert^2 \leq 10^{-4}.
    \label{eq:displacement}
\end{equation}

The optimization procedure will converge once the constraint~(\ref{eq:displacement}) has been satisfied and the improvement in $\varphi_{\mu\nu}(\tau_g)$ between successive iterations drops below a set threshold, chosen to be $10^{-4}$. A maximally entangling gate may be successfully achieved if \mbox{$\varphi_{\mu\nu}(\tau_g) \geq \pi/8$}. For $\Phi\text{M}$ sequences that exceed this value, the Rabi frequency may be scaled down to exactly achieve the target phase $\varphi_{\mu\nu}(\tau_g) = \pi/8$.

\section*{Appendix F: Derivation of filter function for detuning noise}
The filter function framework presented in \cite{Green:2013} and experimentally validated in \cite{Soare:2014} captures the sensitivity of an operator to time-varying noise processes. The expected infidelity of an operation due to a noise process $\epsilon$ can be calculated from the overlap of the noise spectrum, $S_{\epsilon}(\omega)$ and the operation's filter function $F_{\epsilon}(\omega)$,
\begin{equation}
    \label{eq:filter_function}
    \mathcal{I}_\textrm{av} = \frac{1}{2\pi} \int_{-\infty}^\infty d\omega S_{\epsilon}(\omega) F_{\epsilon}(\omega).
\end{equation}
The fidelity of the MS gate is captured by two quantities: (1) the residual qubit-oscillator coupling, and (2) the qubit-qubit entangling phase, $\varphi_{\mu\nu}(\tau_g)$. The former should ideally be zero at the gate's conclusion, whilst maximal qubit-qubit entanglement necessitates that the latter be  \mbox{$\varphi(\tau_g) = \pi/8$}. Here, we modify the filter function framework to derive a filter function $F_{\delta}(\omega)$ that predicts infidelity solely due to residual qubit-oscillator coupling caused by noise on the mode frequencies (a time-dependent detuning error). As the residual qubit-oscillator coupling will be independent of any entangling phase acquired with two or more ions, we proceed by considering a gate-equivalent operation performed on a single ion.     

If the residual motional displacement is small, that is {$\alphaScaledSumSq{\tau_g} \ll 1$}, then from \eqref{eq:one_ion_populations} we see that it can be directly inferred from a measurement of $P_1$,
\begin{align}
    P_1 &\approx 2\alphaScaledSumSq{\tau_g}(\bar{n}_k + 1/2) \nonumber \\
    &\eqdef \sum_k T_k \left\vert f_k^{(1)}\alpha_k(\tau_g)\right\vert^2.
\end{align}
We model trap frequency noise by modifying the laser detuning as \mbox{$\delta_k \rightarrow \delta_k + \epsilon(t)$}, where $\epsilon(t)$ is a zero-mean noise process, altering the phase space trajectories to
\begin{align}
	\alpha_k(\tau_g) &= \Omega\int_0^{\tau_g} dt e^{-i[(\delta_k  + \epsilon(t))t + \phi(t)]}.
	\label{eq:noisy_trajectory}
\end{align}
The assumption that the residual motional displacement remains small requires `weak' noise, that is \mbox{$ \E{\epsilon(t)^2} {\tau_g}^2 \ll 1$}. In experiment, we can only consider noise-averaged measurements, thus we take the ensemble average over this noise process, yielding 
\begin{widetext}
\begin{align}
	\mathbb{E}[P_1] &\approx \E{\sum_k T_k \left\vert f_k^{(1)}\alpha_k(\tau_g)\right\vert^2} \nonumber \\
	&\approx \sum_k T_k\left\vert \Omega \f{1}{k} \right\vert^2  \int_0^{\tau_g} dt_1 \int_0^{\tau_g} dt_2 e^{-i[\delta_k (t_1 - t_2) + \phi(t_1) - \phi(t_2)]}  \E{\left(1 - i\epsilon(t_1)t_1 - \frac{\epsilon(t_1)^2 t_1^2}{2} \right) \left(1 + i\epsilon(t_2)t_2 - \frac{\epsilon(t_2)^2 t_2^2}{2} \right)} \nonumber \\
	&= \sum_k T_k\left\vert \Omega \f{1}{k} \right\vert^2 \int_0^{\tau_g} dt_1 \int_0^{\tau_g} dt_2 e^{-i[\delta_k (t_1 - t_2) + \phi(t_1) - \phi(t_2)]}  \E{\epsilon(t_1)\epsilon(t_2)}t_1 t_2
\end{align}
\end{widetext}
where to arrive at the second line we only consider terms up to \mbox{$\E{\epsilon(t)^2}t^2 \leq  \E{\epsilon(t)^2} {\tau_g}^2$}. Assuming that the gate has been constructed to decouple from all modes in the absence of noise, we simplify further by ignoring all terms dependent on only one integration variable, as they will be multiplied by exactly zero due to the complete mode decoupling condition in the other integral.
The Wiener-–Khinchin Theorem then relates the autocorrelation function in the noise ensemble expectation to the noise spectrum in the frequency domain,
\begin{equation}
    \label{eq:Wiener_Khinchin_Theorem}
    \E{\epsilon(t_1)\epsilon(t_2)} = \frac{1}{2\pi} \int_{-\infty}^\infty d\omega S_\delta(\omega) e^{i\omega(t_1 - t_2)}.
\end{equation}
Applying this, we can rewrite the expectation of $P_1$ in terms of the noise spectrum $S_{\delta}(\omega)$ and the filter function, $F_{\delta}(\omega)$:
\begin{widetext}
\begin{align}
	\mathbb{E}[P_1] 
	&\approx  \frac{1}{2\pi} \int_{-\infty}^\infty d\omega S_\delta(\omega) \sum_k  T_k \left\vert \Omega \f{1}{k} \right\vert^2  \int_0^{\tau_g} dt_1 \int_0^{\tau_g} dt_2   e^{-i[(\delta_k - \omega) (t_1 - t_2) + \phi(t_1) - \phi(t_2)]} t_1 t_2 \nonumber \\
	&= \frac{1}{2\pi} \int_{-\infty}^\infty d\omega S_\delta(\omega)
	\sum_k T_k \left\vert \Omega \f{1}{k} \int_0^{\tau_g} dt e^{-i[(\delta_k  - \omega)t + \phi(t)]} t \right\vert^2 \nonumber \\
	&\eqdef \frac{1}{2\pi} \int_{-\infty}^\infty d\omega S_\delta(\omega) \sum_k F_{\delta,k}(\omega) \nonumber \\
	&\eqdef \frac{1}{2\pi} \int_{-\infty}^\infty d\omega S_\delta(\omega)F_{\delta}(\omega)
	\label{eq:decoupling_filter_function}.
\end{align}
\end{widetext}
The total filter function $F_{\delta}(\omega)$ is defined as a summation of the individual `modal filter functions', $F_{\delta,k}(\omega)$. This spectral overlap predicts the expected value of $P_1$ (and thus the residual motional displacement) in the presence of a given noise spectrum $S_\delta(\omega)$, allowing us to compare the performance of $\Phi$M gates with different levels of noise suppression for each motional mode.

Following a similar procedure, we can also derive $F_{\delta}(\omega)$ for a targeted entangling operation between a pair of ions $\mu,\nu$ in an $N$ ion chain: 

\begin{widetext}
\begin{equation}
F_{\delta}(\omega) = \sum_k T_k\left(\left\vert f_{k}^{(\mu)}\right\vert^2+\left\vert f_{k}^{(\nu)}\right\vert^2\right) \left\vert \Omega \int_0^{\tau_g} dt e^{-i[(\delta_k  - \omega)t + \phi(t)]} t \right\vert^2.   
\end{equation}
\end{widetext}

\null%

\end{document}